\documentclass[12pt,titlepage]{article}
\usepackage{ragged2e}
\usepackage{amsfonts,amscd}
\usepackage{amssymb}
\usepackage{mathtools}
\usepackage{amsmath}
\usepackage{amscd}
\usepackage{lineno,hyperref}
\usepackage{float}
\usepackage{xcolor}
\usepackage[]{graphicx}
\usepackage{subfigure} 
\usepackage{chngcntr}
\usepackage{extarrows}
\usepackage{commath}
\usepackage{comment}
\usepackage{geometry}
\title{Adaptive transitions in FitzHugh-Nagumo networks with Hebb-Oja coupling rules}

\author{A. Provata \thanks{Corresponding author; E-mail: a.provata@inn.demokritos.} \\
\small Institute of Nanoscience and Nanotechnology, \\
\small National Center for Scientific Research “Demokritos”, Athens 15341, Greece \\
\\ 
 G. C. Boulougouris \thanks{E-mail: gbouloug@mbg.duth.gr} \\
\small Institute of Nanoscience and Nanotechnology, \\
\small National Center for Scientific Research “Demokritos”, Athens 15341, Greece  and\\
\small Department of Molecular Biology and Genetics,\\ 
\small Democritus University of Thrace, Alexandroupolis 68100, Greece \\
\\
J. Hizanidis \thanks{E-mail: hizanidis@physics.uoc.gr}\\
\small Institute of Nanoscience and Nanotechnology,\\
\small  National Center for Scientific Research “Demokritos”, Athens 15341, Greece  and\\
\small Institute of Electronic Structure and Laser, \\
\small Foundation for Research and Technology-Hellas, Herakleio, 70013, Crete, Greece}

\date{\today}

\begin{document}
\maketitle

\begin{abstract}
Adaptive coupling in networks of interacting neurons has gained recent attention due to the many applications
both in biological and in artificial neural networks, where adaptive coupling 
or synaptic plasticity is considered as a key factor in
learning processes. In the present study, we apply adaptive connectivity rules in networks of interacting
FitzHugh-Nagumo oscillators. Adaptive coupling, here, is realized via Hebbian learning adjusted by the Oja rule to 
prevent the network link weights from growing without bounds. Numerical investigations demonstrate that during the
adaptation process the FitzHugh-Nagumo network undergoes adaptive transitions realizing traveling waves,
synchronized states and chimera
states transiting through various multiplicities.  
These transitions become more evident when the time scales governing the coupling dynamics are
much slower than the ones governing the nodal dynamics (nodal potentials). 
Namely, when the coupling time scales are
slow, the network has the time to realize and demonstrate
 different synchronization regimes before reaching the final steady
state. The transitions can be observed not only in the spacetime plots
but also in the abrupt changes of the average coupling weights as the network evolves in time.
Regarding the asymptotic coupling distributions, we show that the limiting average coupling strength 
follows an inverse power law with respect to the Oja parameter (also called "forgetting" parameter)
which balances the learning growth. We also report abrupt transitions 
in the asymptotic coupling strengths when the parameter related to adaptive coupling
crosses from fast to slow time scales. 
These findings are in line with previous studies on spiking neural networks. 
\end{abstract}

\justifying
\section{Introduction}
\label{sec:intro}

 During the early stages in the development of network science,
complex networks were generally regarded as static, meaning that their inter-element connectivity remains 
constant in space and time (static networks). 
The complexity of these networks was largely associated with the statistical characteristics of their coupling 
weights \cite{newman:2010,albert:2002,boccaletti:2006,barabasi:2016}.
The dynamics evolving on the often heterogeneous static networks was largely influenced by the intrinsic 
structural complexity of the network, leading to emergent phenomena, such as traveling waves, solitons, 
Turing patterns and
hybrid synchronization patterns \cite{nicolis:2012,kartashov:2011,majhi:2019,zakharova:2020,schoell:2016}.
Inspired by the structural plasticity observed in biological networks (e.g., gene networks, brain networks) 
as well as by the rapid transformations characteristic of social and technological networks, 
it now becomes evident that the attention must be
shifted toward studying the dynamics on temporal and adaptive networks, in which the underlying topology 
evolves concurrently with the system's dynamics. Along these lines, the aim of the present study 
is to explore the influence
of the network adaptivity on the formation and evolution of hybrid synchronization states.
Hybrid synchronization
states, known as chimera states, are observed
in static networks composed by identical and identically linked nonlinear oscillators 
and they are mainly discussed in relation to brain activity 
\cite{haddad:2014,olmi:2019,lazarevich:2017,koulierakis:2020,gerster:2020,chouzouris:2021,ramlow:2019,rattenborg:2006,santos:2017,andrzejak:2016}.
To this purpose, we employ a network composed by FitzHugh-Nagumo oscillators
with adaptive coupling and we explore its response under variations of different model/network parameters.
 The results show that for slow varying coupling strengths the network may undergo transitions between
various states with different synchronization patterns before reaching its asymptotic state.

Adaptivity of the couplings in a network was first addressed in the neuropsychology of 
the brain and is described in
the seminal work ``The organization of behavior: A neuropsychological theory'' by D.~O.~Hebb \cite{hebb:1949}. 
There, he postulated the Hebbian learning principle that ``neurons that fire together wire together''.
The interpretation of this
idea is that the neuronal synapses are not static in time but they may strengthen
 or weaken depending on the local potential
environment.
A first quantitative description of this principle is that the change in the coupling between two neurons is proportional to
the product of the potentials of the pre- and post-synaptic neurons. Numerical investigations using this direct 
quantitative description lead to perpetual increase of the coupling weights (synapses' strengths).
To cure the coupling divergence E. Oja
proposed in 1982 a modification to the quantitative description by introducing 
a  forgetting term, which counteracts to the
product of the pre- and post-synaptic potentials, leading to finite asymptotic couplings \cite{oja:1982,oja:1989}.
The ideas of network adaptivity find now applications in the simulation of networks of interacting
neurons with regards to brain functions such as short-term plasticity, long-term depression, 
long-term potentiation and
structural plasticity \cite{gerstner:2002,morrison:2008, citri:2008, caroni:2012}.
Besides biological, physiological and medical applications, the concept of adaptivity in networks 
has recently been widely applied in the domains of machine learning and artificial intelligence 
\cite{hopfield:1982,carleo:2019}.

Hybrid synchronization states in networks of coupled oscillators
have been studied during the past twenty years, since their first discovery in 2002
by Y.~Kuramoto and D.~Battogtokh \cite{kuramoto:2002,kuramoto:2002a}. The most prominent of these states are the
chimera states, which are characterized by coexistence of alternating coherent and incoherent domains 
\cite{kuramoto:2002,abrams:2004}. While the original studies used the phase oscillator as the coupled unit,
later on, chimera states were discovered for a variety of nonlinear oscillators in 1D ring network geometries, 
such as integrate-and-fire models \cite{tsigkri:2016,tsigkri:2017,parastesh:2021,avitabile:2023,provata:2025b,oomelchenko:2024},
the Hindmarsh-Rose model
 \cite{hizanidis:2014,wei:2018,majhi:2019,kumarpal:2024,majhi:2025}, 
the van der Pol oscillator \cite{ulonska:2016,nganso:2023} and the Stuart-Landau oscillator 
\cite{zakharova:2020,bera:2017}
among many others. Chimera states are also reported in networks of higher spatial dimensions 
\cite{oomelchenko:2012,maistrenko:2015,maistrenko:2017,schmidt:2017,kasimatis:2018,provata:2025}
and higher complexities \cite{kumarpal:2024,majhi:2025,parastesh:2025}.
For the FitzHugh-Nagumo model, in particular,  chimera states were reported on a 1D ring 
\cite{majhi:2019,majhi:2025,omelchenko:2013, omelchenko:2015} 
and 2D torus geometries \cite{koulierakis:2020,schmidt:2017}
giving rise to a variety of chimera patterns such as single- and multi-chimera states, traveling and stable chimeras,
spot and multi-spot chimeras, stripes and rings. 
 All the aforementioned studies concern the case of networks with time independent couplings.
One of the main conclusions of these studies was that, in absence of adaptivity,
 the type of chimera patterns and the number of coherent/incoherent
regions were mostly dependent on the coupling strength and coupling range. The question of including 
coupling adaptivity
and its effects on chimera patterns  arises, thus, naturally in networks of neuronal oscillators.

\par In a previous study, the effects of Hebb-Oja adaptivity were studied in the case of the Leaky Integrate-and-Fire
model \cite{provata:2025c}. There, it was shown that adaptive linking leads to transitions between different 
chimera and bump states as the coupling crosses from negative (inhibitory) to positive (excitatory) values, and the opposite.
In  \cite{provata:2025c}, the role of the coupling time scale parameter was first highlighted and it was shown that,
when the coupling time scales are much larger than the membrane potential time scales, adaptivity drives the system
to cross different synchronization regimes. To the contrary, for similar time scales in the two processes the network
reaches the asymptotic state very fast, without explicit demonstration of intermediate synchronization regimes.
 The present study
tests these ideas on networks of FitzHugh-Nagumo oscillators, which, in conditions of constant (non-adaptive)
 couplings, are known to lead to the formation of distinctive 
chimera states of various multiplicities depending on the network parameters \cite{omelchenko:2013}. In the next sections,
it is numerically shown that Hebb-Oja adaptivity also drives the FHN network to cross regimes of different chimera multiplicity,
provided that the time scale of the coupling adaptivity is much larger than both time scales governing the membrane potential
as well as the recovery potential dynamics. In the sequel, it is also stressed that transitions to bump states are not
observed in the FHN network, since the original, non-adaptive FHN network does not demonstrate bump state dynamics. 

\par The present study is organized as follows. In the next section, Sec.~\ref{sec:model}, the FHN network model is presented
with adaptive coupling following the Hebb-Oja rule. A presentation of the  quantitative methods employed 
is included in the same section. Such methods comprise the Kuramoto order parameter, 
the average effective coupling and its spatial and temporal fluctuations.
In Sec.~\ref{sec:results}, the transitions between states with different
synchronization properties are illustrated using spacetime plots. Quantitative results regarding the
detection of the synchronization transitions
are presented in the same section. In particular, in Secs.~\ref{sec:results-1} and~\ref{sec:results-2}
the cases of increasing coupling strengths are discussed, while the case of decreasing coupling  
 is presented in
Sec.~\ref{sec:results-3}. The influence of the adaptivity parameters (Oja coupling parameter $\alpha$ and
 and time scale coupling parameter $\tau_{\sigma}$) is
discussed in Sec.~\ref{sec:results-4}.
In the Conclusions, the main findings of this work are recapitulated
and open problems for further studies are proposed. 

\section{Model and Methods}
\label{sec:model}
 The FitzHugh-Nagumo (FHN) model has its origins in the Hodgkin-Huxley model and is designed to capture 
the most essential dynamics of the activity of a single neuron cell \cite{hodgkin:1952,fitzhugh:1961}.
Large cohorts of these interconnected neurons are now employed to serve as models of
the brain and other complex dynamical systems. During the past twenty years, the need to address the dynamics of  
such large systems has led to the design of efficient algorithms to simulate networks
with intricate connectivities and nonlinear neuronal dynamics imitating the neuronal activity.
And while the network connectivities are mainly dictated by the physical system (e.g., the brain),
for the nodal dynamics a number of nonlinear neuron models have been employed, 
such as the Integrate-and-Fire \cite{lapicque:1907}, FitzHugh-Nagumo \cite{fitzhugh:1961},
Izhikevich \cite{izhikevich:2001,izhikevich:2007}, Hindmarsh-Rose \cite{hindmarsh:1982,hindmarsh:1984}
 and many others. These models, although they are not as complete as the Hodgkin-Huxley model,
 retain the most essential features of the neuronal activity while they simplify the simulations in terms of
algorithmic complexity and reduced simulation time. One more feature of the networks which will be used in
the current study concerns the adaptivity/plasticity of the link weights between the different neurons of the network,
which finds its origin in the earlier works of D. O. Hebb \cite{hebb:1949} with later modifications
by E. Oja \cite{oja:1982,oja:1989}. 
\par In the next three subsections the main models and methods used in the present study are described. First, 
in Sec.~\ref{sec:network}, the equations describing the evolution of the adaptive FHN network are introduced,
 in Sec.~\ref{sec:quantitative}, the various measures needed for the quantification of the system dynamics
are briefly recapitulated and, in Sec.~\ref{sec:parameters}, a detailed description
of the working parameter set is provided.

\subsection{Networks of FitzHugh-Nagumo oscillators with adaptive coupling}
\label{sec:network}
To study the effects of adaptive coupling on the synchronization in networks of FHN elements,
we consider the simplest connectivity architecture:
a ring network composed of $N$ FHN oscillators with periodic boundary conditions. 
Each FHN oscillator
is situated on the nodes $j=1, \cdots ,N$
of the network and its state is described by two potential variables: $u_j(t)$, 
 which 
accounts for a nonlinear increase of the neuron membrane potential,
and $v_j(t)$ which is a linear recovery potential.\footnote {For brain neurons, 
the membrane (action) potential, $u_j$, 
represents the phase where voltage-gated sodium channels open, 
causing positively charged Na$^+$ ions to rush into the cell 
and the membrane potential increases toward positive values. The recovery potential,
$v_i$, represents activation of the potassium channels, 
allowing K$^+$ ions to exit the membrane and helps to restore the
steady state membrane potential to negative values.} 
The network connectivity is further described by the
variables $\sigma_{jk}(t)$, representing the time dependent weights of the connectivity 
between pre-synaptic node $k$ and post-synaptic neuron $j$.
The equations representing the FHN ring network are:
\begin{subequations}
\begin{flalign}
\label{eq01a} 
\begin{aligned}
 \epsilon \frac{du_{j}(t)}{dt} &= u_{j}(t)-\frac{u_j^3(t)}{3}-v_j(t)
 +  \\ &+  \frac{\sigma_c}{2R}
  \sum_{k=j-R }^{j+R} \sigma_{jk}(t) \Biggl[
b_{uu} \left[ u_{k}(t) - u_{j}(t)\right] +b_{uv} \left[ v_{k}(t) - v_{j}(t)\right] \Biggr] 
\end{aligned}
\end{flalign}
\begin{equation}
\begin{aligned}
 \frac{dv_{j}(t)}{dt} &=u_{j}(t)+\gamma + \\
&+\frac{\sigma_c}{2R} \sum_{k=j-R }^{j+R}  \sigma_{jk}(t)\Biggl[
b_{vu} \left[ u_{k}(t) - u_{j}(t)\right] +b_{vv} \left[ v_{k}(t) - v_{j}(t)\right] \Biggr]
\end{aligned}
\label{eq01b}
\end{equation}
\begin{equation}
\tau_{\sigma}\frac{d\sigma_{jk}(t)}{dt}=u_j u_k -\alpha u_j u_j \sigma_{jk}. \hskip 6cm
\label{eq01c}
\end{equation}
\label{eq01}
\end{subequations}
Different types of parameters are used in Eqs.~\eqref{eq01}. Parameters $\epsilon$ and $\gamma$
characterize the properties of the nodal dynamics and are the same for all elements (nodes).
Parameters $R$, $\sigma_c$, $b_{uu}$, $b_{vu}$, $b_{uv}$ and $b_{vv}$ describe the network
connectivity and are also common to all links. Parameters $\tau_c$ and $\alpha$ control 
the evolution of the connectivity variables $\sigma_{jk}$.
The parameter interpretation is as follows: 
\begin{enumerate}
\item $\epsilon$ is a parameter which 
defines the time scale difference between the membrane and recovery potentials. It is
identical for all FHN elements.
\item
Parameter $\gamma$ controls the stability and type of the system's steady state. 
For uncoupled FHN neurons, if $|\gamma | < 1$ the neuron
 potential performs periodic oscillations. In the present study, we  use 
$\gamma =0.5$, setting the neurons into the oscillatory regime. It is also
considered identical for all neurons.
\item $R$ is the coupling range in the network. Each node is connected with $R$
elements (neighbors) to its left and $R$ elements to its right, on the ring. Periodic 
boundary conditions apply to ensure that all elements have the same number of
connections (identical connectivity). 
\item Parameter $\sigma_c$ controls the sign of the connectivity term, 
allowing it to acquire positive (excitatory) as well
as negative (inhibitory) values.
\item Regarding the full strength of the connectivity , during the system integration both values 
$\sigma_c$ and $\sigma_{jk}$ contribute to the coupling weight
between nodes $j$ and $k$. Their combined effect is called ``effective coupling weight'' 
and is denoted as $\sigma^{\rm eff}_{jk}(t)=\sigma_c \sigma_{jk}(t)$.
\item The parameters $b_{uu}$, $b_{vu}$, $b_{uv}$ and $b_{vv}$ constitute a 2-dimensional
square matrix, $B$,  allowing for cross-coupling interactions between membrane and recovery
potentials, $u$ and $v$, respectively.
\item Parameter $\tau_{\sigma}$  allows to select different time scales for the evolution of the potential
variables, $u_j(t)$ and $v_j(t)$, and for the coupling strength ones, $\sigma_{jk}(t)$. 
\item Finally, the parameter $\alpha$ has been introduced by E. Oja \cite{oja:1982,oja:1989} to balance the
Hebbian learning term in Eq.~\eqref{eq01c} and is further described in the sequel.
\end{enumerate}
\par In the absence of the coupling terms in Eqs.~\eqref{eq01a} and \eqref{eq01b} the elements are 
independent and, for the parameter values selected in this study ($\epsilon=0.01$, $\gamma=0.5$),
they perform free nonlinear oscillations.  Note that in the absence of the coupling control,
e.g. for $\sigma_c=0$, Eq.~\eqref{eq01c} becomes obsolete.
\par If adaptivity is not considered, 
then $\sigma_{jk}$ become time independent and the coupling weights keep their
initial values throughout the simulations. 
For the case of constant (non-adaptive) and identical $\sigma_{jk}=\sigma$ for all nodes,
chimera states of various multiplicities have been observed depending on the connectivity range $R$
and the (constant and identical) coupling strengths $\sigma$
 \cite{omelchenko:2013,omelchenko:2015,schmidt:2017}.
\par As stated earlier, the introduction of adaptivity in the coupling weights is motivated 
by brain dynamics and the
Hebbian principle of adaptivity \cite{hebb:1949}. In Eq.~\eqref{eq01c},  the Hebbian learning
is represented by the first term on the right-hand side (rhs), proportional to $u_ju_k$.
The second term on the rhs in the same equation is the forgetting term, as introduced by E. Oja
to counterbalance the unlimited increase of the coupling weights induced by the
first term. 
The balance between the two terms allows to settle the coupling weights to steady state values which depend only
on the Oja parameter $\alpha$.
Note that $\sigma_{jk}(t)$ represents the coupling from pre-synaptic neuron $k$ to post synaptic
neuron $j$. Therefore, the forgetting term is proportional to the load/charge of the post-synaptic neuron.


\subsection{Quantitative Description}
\label{sec:quantitative}

Because of the adaptivity rules, the various measures quantifying synchronization and complexity
in the system/network are time dependent and only at the asymptotic limit they acquire constant values,
provided that such an asymptotic limit exists.
For quantifying the results of the simulations and assessing the complexity of the produced
synchronization patterns we use the following measures: a) the Kuramoto order
parameter, $z$, b) the average effective coupling strength, $\sigma^{\rm eff}$,
and c) the deviation of $\sigma^{\rm eff}$ around the mean, $D_{\sigma}(t)$.
\par The average effective coupling strength in the system is time dependent and is computed as:
\begin{eqnarray}
\left< \sigma^{\rm eff}(t)\right>=\frac{\sigma_c}{2 R N} \sum_{j=1}^N\sum_{\substack {k=1\\ k\ne j}}^N\sigma_{jk}(t) .
\label{eq02}
\end{eqnarray}
\par The spatial deviation around the mean, $D_{\sigma}(t)$, also varies with time
and shows the instantaneous spreading of the coupling values around the average. Its squared value
is defined as:
\begin{eqnarray}
D^2_{\sigma}(t)=\frac{1}{2 R N} 
\sum_{j=1}^N\sum_{\substack {k=1\\ k\ne j}}^N\Big( \sigma_c\sigma_{jk}(t)-\left< \sigma^{\rm eff}(t)\right> \Big)^2 .
\label{eq03}
\end{eqnarray}
\noindent Because each of the $N$ nodes maintains non-zero weighted connections with only $2R$ other nodes in the network,
the averages in Eqs.~\eqref{eq02} and~\eqref{eq03} are calculated using the $2 R N$
active (non-zero) connections, instead of all $N(N-1)$ potential links, many of which have zero weight.
Therefore, the zero-weighted connections are excluded by construction from the above sums/averages.

\par The Kuramoto order parameter, $z$, was proposed to quantify the degree of synchronization in a network and, for
its definition, the phase variable $\theta_j$ of oscillator $j$ needs to be defined first \cite{kuramoto:2002}. 
For the FHN oscillators
the phase variables are time dependent and  are defined as : \footnote{ For a more formal definition 
 of more broad validity, especially for the cases where noise is included in the dynamics, the so-called
Hilbert transform is used \cite{pikovsky:2001}.}
\begin{eqnarray}
\theta_j(t)=\arctan \Big( \frac{v_j(t)}{u_j(t)} \Big) .
\label{eq13}
\end{eqnarray}
Using this phase definition, the time dependent Kuramoto order parameter, $z(t)$, is determined as 
\cite{kuramoto:2002,pikovsky:2001} :
\begin{eqnarray} 
& z(t)=\big| Z(t) \big| = \frac{1}{N} \left|  \sum\limits_{j=1}^{N}{\rm e}^{i \, \theta_j} \right| ,
\label{eq14} 
\end{eqnarray}
\noindent where the sum runs over all $N$ elements in the network. $z$ takes its minimum value, $z=0$,
when the system is in full asynchrony and all phases are randomly distributed between 0 and $2\pi$.
On the other end, the $z$ maximum value is 1 and is achieved when all elements are fully synchronous. 
At intermediate  values, $0 < z < 1$, the network demonstrates mixed behavior, contains both synchronous
and asynchronous oscillators and can host chimera and/or bump states. Therefore, in systems of coupled
oscillators, intermediate $z$-values (above 0 and below 1) indicate the presence of hybrid states. 

\par Other measures, such as the mean phase velocities that are 
usually employed to characterize the synchrony is networks of
coupled oscillatory units, are not useful in this study because they rely on long-time averages for reliable
estimations. However, in  the present system the continuous evolution of the coupling strengths induces instantaneous
variations in the mean phase velocities and this renders their  numerical estimation 
unreliable.
\par As briefly mentioned in the previous paragraphs, the  coupling strengths start from random initial conditions
and they slowly evolve toward their steady state values. The mean
field asymptotic values of the coupling strengths $\sigma_{jk}$ result
from setting  the right hand side of Eq.~\eqref{eq01c} to zero. This leads to \footnote{ This asymptotic state is
realized in the cases where, at the steady state, the coupling weight distributions are such that the correlation condition
$\left< u_j u_k\right> = \left< u_j u_j\right>$, $\forall (j, k) $ is satisfied \cite{oja:1982,oja:1989}. The 
fulfillment of this correlation condition can be checked numerically.}
\begin{eqnarray}
\sigma_{jk}\> \underset{t\to\infty}{\longrightarrow} \> \begin{cases} 1/\alpha & \forall \> j \ne k \\ 0 & \forall \> j =k  \end{cases} ,
\label{eq06}
\end{eqnarray}
and consequently
\begin{eqnarray}
\sigma^{\rm eff}_{jk}\> \underset{t\to\infty}{\longrightarrow} \> \begin{cases} \sigma_c/\alpha & \forall \> j \ne k \\ 0 & \forall \> j =k  \end{cases} .
\label{eq07}
\end{eqnarray}
\par To achieve maximal diversity in the system and allow the network to pass from as many different synchronization
states as possible, it is advisable to start from initial conditions, $\sigma_{ij}(t=0)$, 
as far as possible from their asymptotic
value $1/\alpha$. E.g., in some cases of the present study, we start from $\sigma_{ij}(t=0)=-0.2$ and,
with Oja parameter $\alpha =1$, the system crosses
into positive couplings ending asymptotically to $\sigma_{ij}\underset{t\to\infty}{\longrightarrow} 1$. 

\subsection{Working Parameters}
\label{sec:parameters}

The present section provides a detailed description of the working parameter set used in this study.
The parameters related to the nodal dynamics are $\epsilon=0.01$, $\gamma=0.5$. Note that the small value
of the parameter $\epsilon$ indicates that the recovery potential $v_j$ evolves in much slower 
time scales than the membrane potential $u_j$.
The parameters related to the network connectivity take indicative values $R=260$ or $R=350$; $\sigma_c=0.2$
for excitatory and $\sigma=-0.2$ for inhibitory interactions. 
The values of the rotational matrix $B$ are defined via a rotation coupling angle $\phi$ as follows: the 
direct connectivity couplings are $b_{uu}=b_{vv}=\cos{\phi}$ and the 
 cross-connectivity ones are $b_{uv}=-b_{vu}=\sin{\phi}$. Previous studies (\cite{omelchenko:2013,omelchenko:2015})
have demonstrated that
$\phi$-values near $\pi /2$ lead to hybrid synchronization (chimera) states and for this reason the 
value $\phi=\pi/2-0.1$ (in rads) is used in the present study.
 Regarding the parameters related to adaptivity, we use $1\le \alpha \le 10$ and $\tau_{\sigma}$ takes values
in long and short scales as : I) $\tau_{\sigma}=1$ for time scales comparable
to the ones of the membrane potential, 
II) $\tau_{\sigma}$=100 for time scales two orders of magnitude slower and, 
III) $\tau_{\sigma}$=1000 for time scales 
three orders of magnitude slower than the ones of the membrane potential. 
The results were also verified for 
larger time scales, $\tau_{\sigma}= 10000$.
For the network integration, the forward Euler method was used with integration time step $dt=0.001$.
Then, 1 Time Unit (TU) corresponds to 1000 integration time steps. 
 For verification purposes, in several cases the 4th order Runge-Kutta method with $dt=0.01$ was employed.
\par Initial conditions (potentials, $u_j(t=0)$) are randomly and homogeneously
distributed in the interval $-2 \le u_j(t=0) \le 2$ and, to facilitate initial settings,
the corresponding recovery potentials are adjusted to $v_j(t=0)=\pm \sqrt{4-u_j^2(t=0)}$. 
Initial coupling weights, $\sigma_{ij}(t=0)$, are selected
 to be homogeneous with values as far away as possible from the inverse Oja parameter.
Homogeneity is not important in the initial couplings,
because immediately after the simulation starts their values spread randomly due to the
interactions with the randomly distributed potential values, see Eq.~\eqref{eq01c}. It is important, though,
to chose initial couplings away from the inverse value of the Oja parameter, 
to allow ample time for the evolution of the couplings
toward their asymptotic values, $\sigma_{ij} \to 1/\alpha$, see Eqs.~\eqref{eq06} and~\eqref{eq07}.

\section{Results}
\label{sec:results}
In the next three sections, three exemplary cases of adaptive transitions in FHN networks with Hebb-Oja rules
are presented. The first and second 
sections concern gradually increasing average couplings, whereas in Sec.~\ref{sec:results-3} the
average coupling is gradually decreasing. Section~\ref{sec:results-4} is devoted to the exploration of
the network activity in relation to the adaptive system parameters $\alpha$ and $\tau_{\sigma}$, and Sec.~\ref{sec:results-5} discusses briefly the idea of using large time scales in the coupling adaptivity. 

\subsection{Increasing average coupling}
\label{sec:results-1}
\begin{figure}[H]
\includegraphics[width=1.0\textwidth]{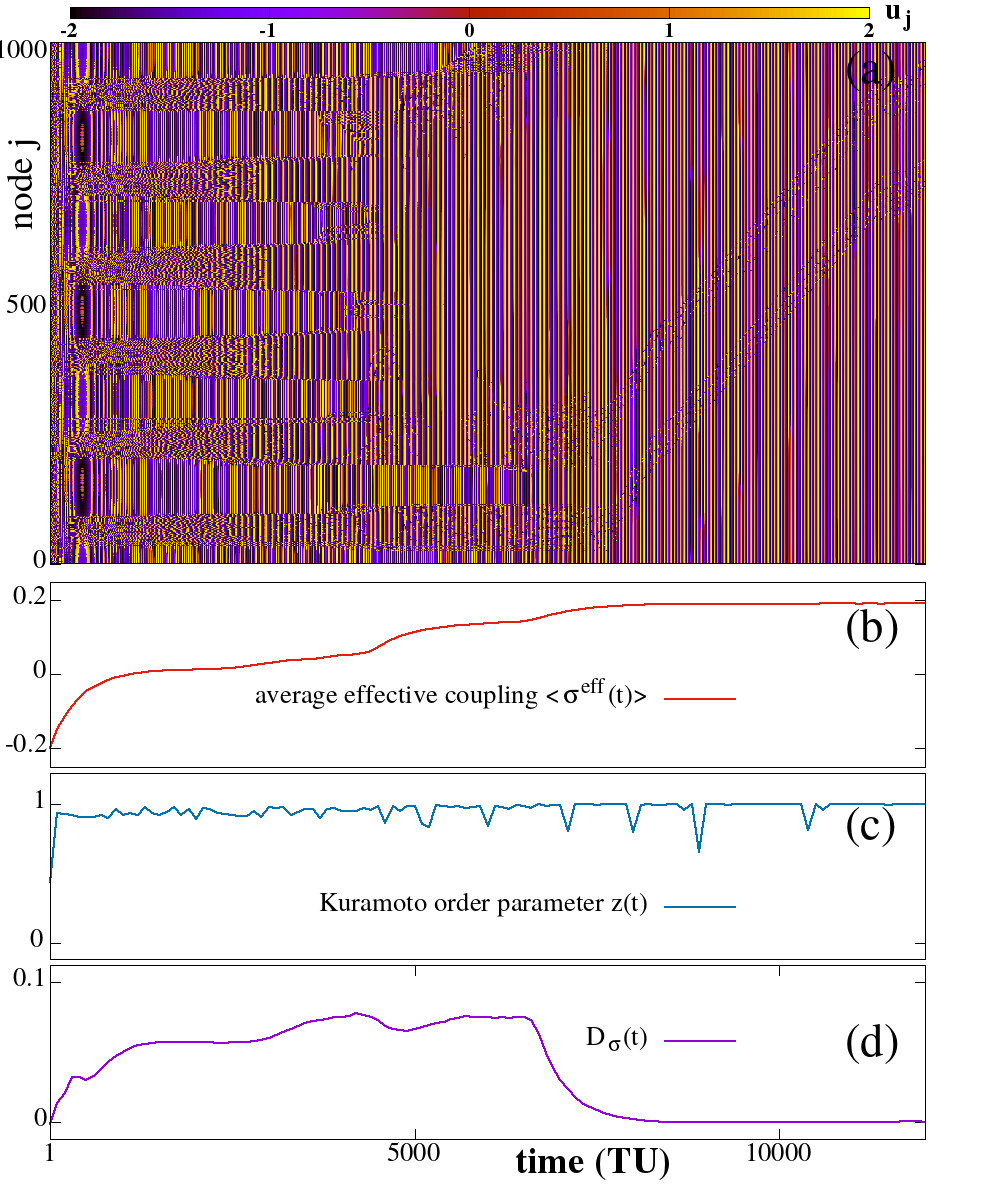}
 \caption{\label{fig01} FHN network with adaptive Hebb-Oja coupling: (a) Spacetime plot; the time evolution is
on the x-axis, the node index, $j=1 \cdots N$, is on the y-axis, and the color represents the state of 
the membrane potential at time $t$, $u_j(t)$.
The different synchronization regimes are clearly visible in the space time plot.
(b) Evolution of the effective average coupling, $\sigma^{\rm eff}(t)$. (c) Evolution of the Kuramoto
order parameter, $z(t)$. (d) Evolution of the coupling fluctuations, $D_{\sigma}(t)$. 
Parameter values are:
$\epsilon=0.01$, $\gamma=0.5$, $R=260$, $\sigma_c=+0.2$,
 $\phi=\pi/2-0.1$ and the integration time step is $dt=0.001$. 
The parameters related to adaptivity are: $\alpha =1$ and $\tau_{\sigma}=1000$. The initial conditions
for the membrane potential were randomly and homogeneously distributed in the range $-2.0 \le u_j(t=0) \le 2$. 
}
\end{figure}

The best way to present the transitions of the network through different synchronization regimes is by using
the spacetime plots. In Fig.~\ref{fig01}(a), we present the spacetime plot for a ring with $N=1024$ FHN oscillators
linked via Hebb-Oja coupling (see Eq.~\eqref{eq01}). The parameters used are described in the working parameter 
set (see Sec.~\ref{sec:parameters}) with $R=260$, $\tau_\sigma = 1000$, $\alpha =1$ and $\sigma_c=+0.2$.
The initial potentials are randomly selected in the interval $-2 \le u_j(t=0) <2$, while $\sigma_{jk}(t=0)=-1$ for 
$ j-R \le k \le j+R $ and 0 otherwise.
Several transition regions may be observed. During early
times, $t < 500$ TU, the network is disorganized due to the random initial potentials. 
Regarding the average effective couplings, at early times they take values $\left<\sigma^{\rm eff}(t=0) \right> =
\sigma_c \cdot \left<\sigma_{ij}(t=0) \right> =-0.2$.
The initial random state of the network is also confirmed in
Fig.~\ref{fig01}(c), where the Kuramoto order parameter $z$ takes small values. 
As time increases, in the interval $500 < t < 3000$ the FHN network develops 6 coherent and 6 incoherent regions,
realizing a 6-headed chimera state. In this region, the average coupling strength increases, see Fig.~\ref{fig01}(b),
while the Kuramoto order parameter reaches high values, but not as high as $z=1$ which corresponds to full
coherence, see Fig.~\ref{fig01}(c). At times $t\sim 3000$ a new transition is observed: the interior of the
incoherent domains reduces to coherence while their borders retain the incoherent state for another 1000 TUs.
This transition is reflected by the evolution of the average coupling strengths, which around $t\sim 3000$ TU shows
a gradual linear increase after a period of constant values, see Fig.~\ref{fig01}(b). In the same region
 the Kuramoto index increases progressively toward unity, as the interiors of the 
incoherent regions tend to synchronize. Around $t \sim 4500$ TU most of 
the incoherent domains vanish and only 2-3 of then survive until $t \sim 7000$ TU as seen in the spacetime plot.
 The transitions are still visible in
Fig.~\ref{fig01}(b), where the effective coupling undergoes abrupt changes. The Kuramoto order parameter
at $t\sim 4500$ TU reaches values close to 1 (almost full coherence) with considerable fluctuations in time.
Finally, around $t \sim 7500$ TU the system undergoes another transition leading to the asymptotic state
which consists of 2  incoherent domains traveling in a sea of coherent elements. In this region, $t> 7500$ TU,
the average effective coupling has reached its asymptotic value, $\left<\sigma^{\rm eff} \right> =
\sigma_c \left<\sigma_{ij} \right> =\sigma_c / \alpha=0.2$, close to the expected mean field approximation,
see Eq.~\eqref{eq07} and Fig.~\ref{fig01}(b).

\par The transition points are also reflected in the evolution of the coupling fluctuations. 
In Fig.~\ref{fig01}(d) we note that the transitions between different chimera states 
are accompanied
by corresponding amplifications of the coupling fluctuations, $D_{\sigma}$. 
This takes place until the final state is reached, at about $t \sim 7500$ TU,
 when most of the system shows coherence with few minimal incoherent domains.

\subsection{Increasing average coupling and effects of initial conditions}
\label{sec:results-2}
\begin{figure}[H]
 \centering
        \includegraphics[width=1.0\textwidth]{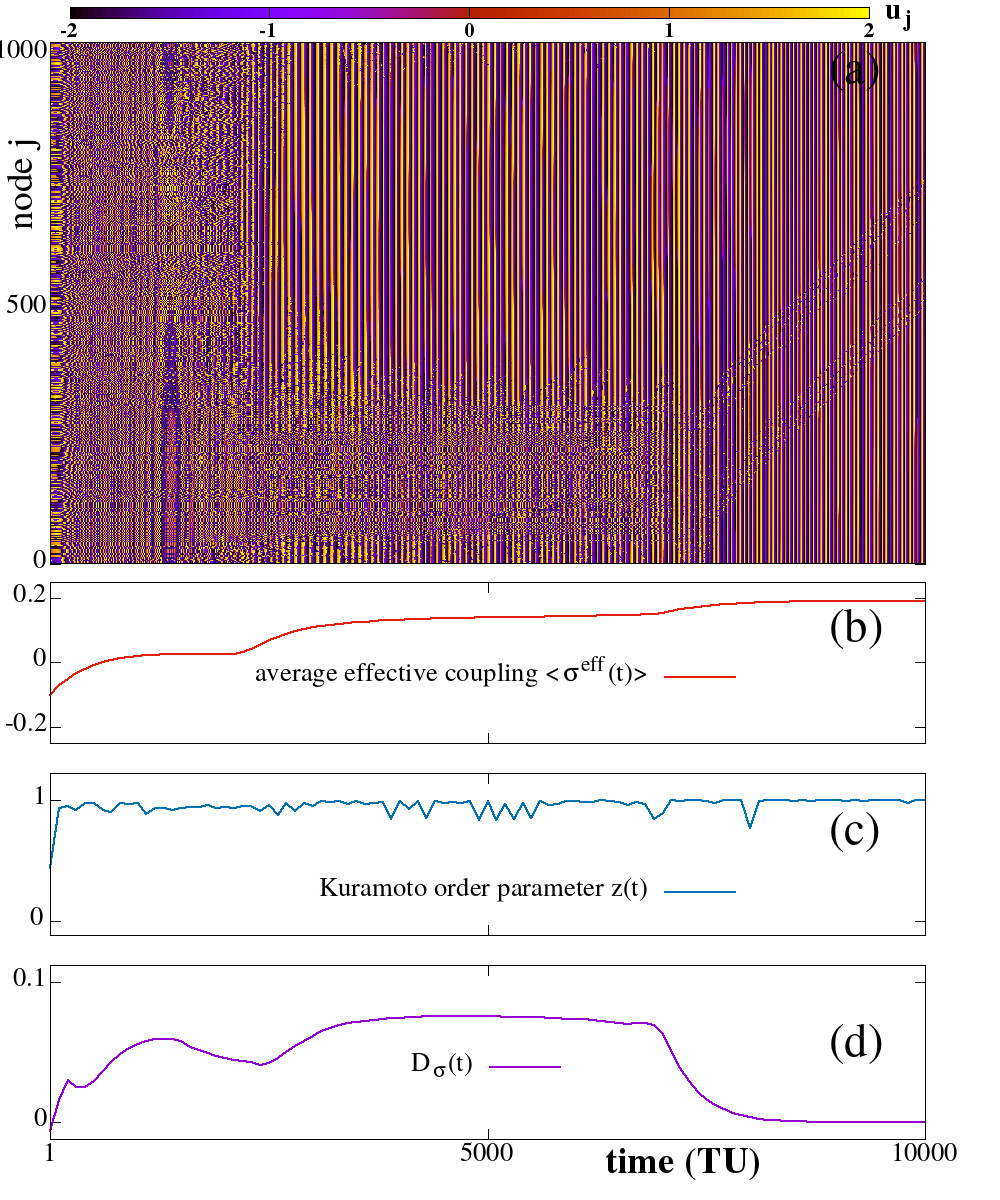}
 \caption{\label{fig02} FHN network with adaptive Hebb-Oja coupling and different
initial conditions with respect to Fig.~\ref{fig01}: (a) Spacetime plot; the time evolution is
on the x-axis, the node index, $j=1 \cdots N$, is on the y-axis, and the color represents the state of 
the membrane potential at time $t$, $u_j(t)$.
The different synchronization regimes are clearly visible in the space time plot.
(b) Evolution of the effective average coupling, $\sigma^{\rm eff}(t)$. (c) Evolution of the Kuramoto
order parameter, $z(t)$.  (d) Evolution of the coupling fluctuations, $D_{\sigma}(t)$.
All parameter values are the same as in Fig.~\ref{fig01} but the initial coupling weights
$\sigma_{jk}(t=0)$ are different in the two figures. 
}
\end{figure}

The previous results show an exemplary case of adaptive transitions. 
Starting from different initial coupling values, now initiated around $\sigma_{jk}(t=0)=-0.5$,
or $\sigma_{jk}^{\rm eff}(t=0)=-0.5\cdot 0.2=-0.1$, a different route to the
asymptotic state may be observed, see Fig.~\ref{fig02}. (All other parameters
stay the same, as in Fig.~\ref{fig01}.)

\par  Figure ~\ref{fig02}(a) presents the spacetime plot of the FHN network as is transiting through 
different synchronization regimes. 
Starting from random initial conditions and low negative coupling strengths 
 (see Fig.~\ref{fig02}(b)),
the network demonstrates asynchronous dynamics as is also testified by the low Kuramoto order parameter
(see Fig.~\ref{fig02}(c)). Very soon, at times as short as 100 TUs, the system enters a
state of incoherence as the coupling strength builds up until about 2500 TUs 
(see Fig.~\ref{fig02}(b)). The increase of the coupling strength is not continuous: in the time interval 
$[1000 - 2000]$ TU the coupling strength seems to stabilize, but just afterwards a new
dynamical regime emerges where the system organizes into a chimera state with one large coherent
and one smaller incoherent domain. This regime lasts until about 7000 TU and then,
after another abrupt coupling increase, the
final state emerges, where the coupling stabilizes at 
around $\sigma^{\rm eff}=+0.2$, close to the mean field expected value  $ {\sigma_c}/\alpha$, 
and the final state consists of two small incoherent domains mediated by one large and one small
coherent domain. At this asymptotic state, the fluctuations $D_\sigma$ of the coupling
strength are consistent with the presence of a chimera state where the hybrid state emerges
even in the case of completely equal coupling strengths.
\par Note that, although the networks have started with different initial conditions
in Figs.~\ref{fig01} and~\ref{fig02}, they end up in asymptotic states with the same
structural characteristics dictated by the equal average final coupling strengths.
\par The two examples demonstrate the sensitive dependence on initial conditions
as reflected in the system dynamics.
Starting from different initial conditions, the system may use different adaptation routes
before reaching its asymptotic state.

\subsection {Decreasing average coupling}
\label{sec:results-3}

\par
\begin{figure}[H]
        \includegraphics[width=1.0\textwidth]{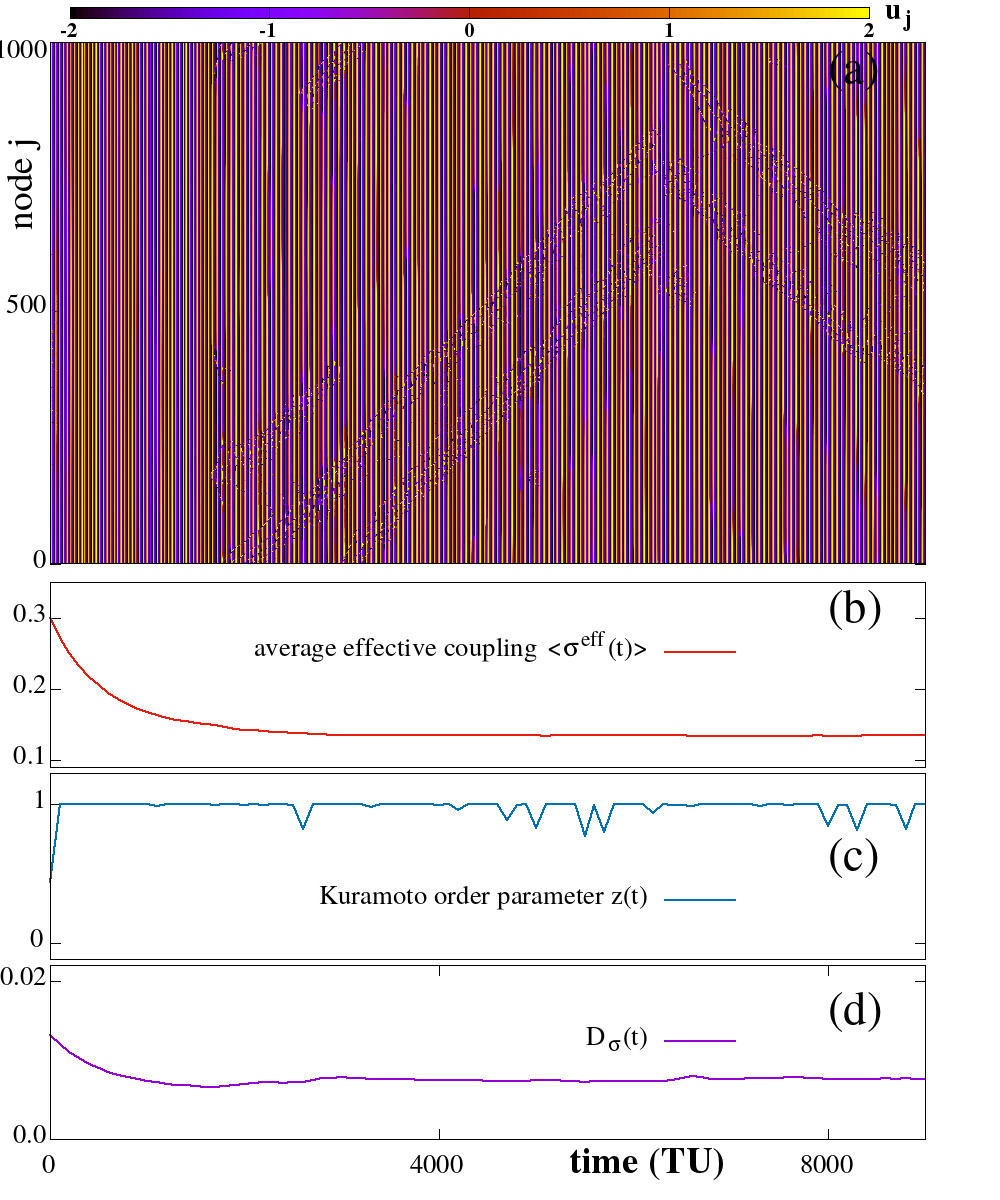}
 \caption{\label{fig03} FHN network with adaptive Hebb-Oja coupling and different
initial conditions with respect to Fig.~\ref{fig01}: (a) Spacetime plot; the time evolution is
on the x-axis, the node index, $j=1 \cdots N$, is on the y-axis, and the color represents the state of 
the membrane potential at time $t$, $u_j(t)$.
The different synchronization regimes are clearly visible in the space time plot.
(b) Evolution of the effective average coupling, $\sigma^{\rm eff}(t)$. (c) Evolution of the Kuramoto
order parameter, $z(t)$. (d) Evolution of the coupling fluctuations, $D_{\sigma}(t)$.
The control parameter is $\sigma_c=+0.1$, the Oja constant is
$\alpha=0.7$ and the average initial coupling 
starts from $\sigma_{ij}^{\rm eff} (t = 0) = 0.3$.
All other parameter values and conditions are the same as in Fig.~\ref{fig01}. 
}
\end{figure}

 The reverse evolutionary process of effective coupling strengths, progressing from high coupling values
to lower ones, 
is illustrated in  Fig.~\ref{fig03}. The system has started from randomly and homogeneously distributed
initial potentials which are hard to discern in Fig.~\ref{fig03}(a) because the network nodes synchronize very fast.
This fast synchronization process is testified by Fig.~\ref{fig03}(c), where the Kuramoto order parameter
shows a fast increase from low values (indicating full asynchrony) toward 1 (compatible with full system synchrony
achieved around $t \sim 100$ TU).
The initial values of coupling strength were set to high values, $\sigma^{\rm eff}_{ij}(t=0)=0.3$,
with Oja constant $\alpha =0.7$. 
\par After the initial state, adaptation drives the network to lower
coupling values, as shown in Fig.~\ref{fig03}(b). After the short initial stage (100 TU),
the system enters a synchronous phase which lasts for about 1500 TU. 
During  this time interval all elements oscillate synchronously. 
This synchronous phase is clearly seen in Fig.~\ref{fig03}(a)
and in Fig.~\ref{fig03}(c), where the synchrony is verified by the Kuramoto order parameter persistently 
staying near $z\sim 1$. During this same time the average effective coupling drops continuously.
Around $t \sim 1500$ TU fully synchrony
 gets destroyed by the coupling fluctuations produced by the Hebb-Oja adaptation
rule and three small incoherent regions emerge. This state survives very shortly and gives rise to
a 2-chimera state consisting of two small incoherent domains separated by a small and a large coherent ones.
The two incoherent domains travel with a constant velocity around the network. The 2-chimera appears
around $t \sim 1700$ TU and persists thereafter. Note that, around $t \sim 6000$ TU, there is a reversion of the
traveling direction of the incoherent regions, but all other characteristics of the system are retained.
During this last phase, the system has reached its asymptotic state and the average effective coupling
reaches the expected value  $\sigma^{\rm eff}_{ij}=\sigma_c/\alpha =0.1/0.7=1.42$.
\par Overall, in this case the coupling adaptivity has led to destabilization of the network synchrony.
We note that even
in the cases where the system has reached full synchronization, tiny fluctuations may destabilize locally
the system giving rise to
chimera states if the parameters are driven to appropriate values.

\subsection {Effects of the adaptivity parameters}
\label{sec:results-4}
We here examine the influence of the adaptivity parameters in the network dynamics, scanning
different values of $\alpha$ and $\tau_{\sigma}$ .
\par
\begin{figure}[H]
 \centering
\includegraphics[width=0.490\textwidth]{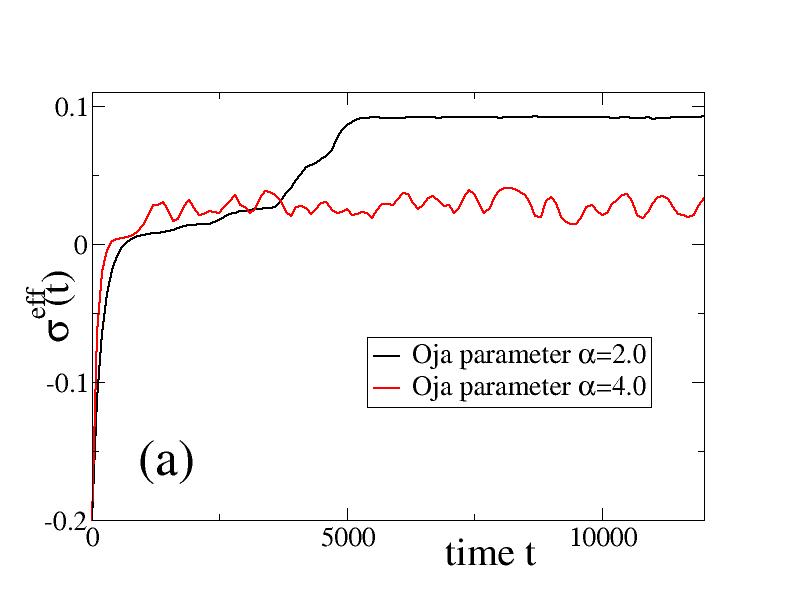}
\includegraphics[width=0.490\textwidth]{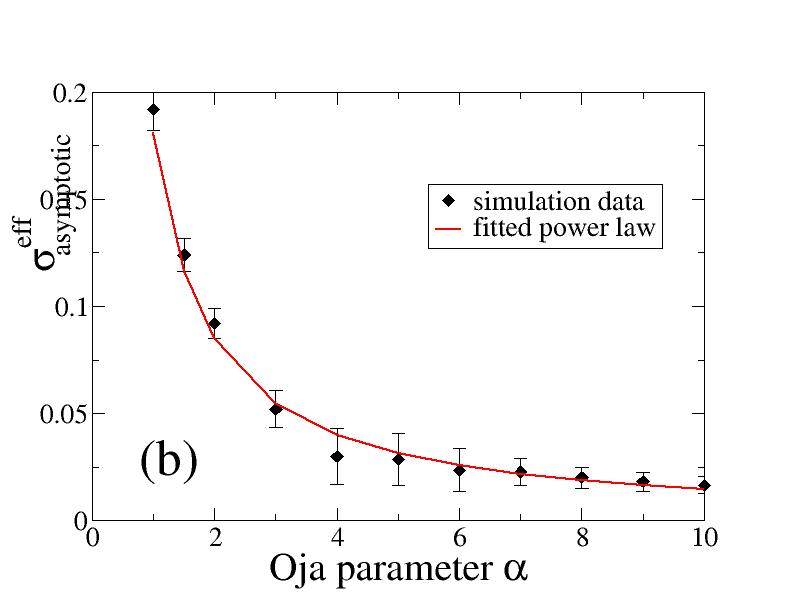}
 \caption{\label{fig04} FHN network with adaptive Hebb-Oja coupling and different Oja
parameters.
 (a) Typical temporal evolution of the coupling
for different values of $\alpha$. In the case $\alpha=2.0$ (black curve) 
the effective coupling goes to the asymptotic state via abrupt transitions, while in case 
$\alpha=0.4$ (red curve) the network reaches rapidly a fluctuating asymptotic state. 
(b) The effective asymptotic coupling as a function of the Oja coupling $\alpha$; 
the black diamonds are simulation results and the red line is a fitted curve. 
All other parameter values are the same as in Fig.~\ref{fig01}. 
}
\end{figure}

 To explore the influence of the Oja parameter $\alpha$ we present, in Fig.~\ref{fig04}(a),
the temporal evolution of the average effective coupling as a function of time for 
 $\alpha =2 $ (black line) and $\alpha =4 $ (red line). Both curves demonstrate variability
in their evolution:  the case $\alpha =2 $ is characterized by successive abrupt
transitions followed by relaxation periods before the asymptotic state is reached,
whereas in the case $\alpha =4 $ the network quickly enters the asymptotic state
demonstrating considerable fluctuations in the effective average coupling.
Other such examples for $\alpha =1$ can be seen in Figs.~\ref{fig01}(b) and~\ref{fig02}(b),
and for $\alpha =0.7$ in Fig.~\ref{fig03}(b). The route to the asymptotics is then governed
by the initial state forcing the network to evolve either smoothly or irregularly
(discontinuously) toward its final regime.
\par Considering the final (asymptotic) state, its characteristics are dictated by the network parameters
$\alpha$ and $\sigma_c$ and is independent of the route the system has followed before 
reaching it. This may be observed in the previous section by comparison of  Figs.~\ref{fig01} and~\ref{fig02}. Regarding the asymptotic state, in Fig.~\ref{fig04}(b) we present the average
coupling $\sigma^{\rm eff}_{\rm asymptotic}$ as a function of $\alpha$ starting from the
same initial conditions, $\sigma_c=+0.2$ and all other parameter 
values as in the working parameter
set. A fitting (red) curve in Fig.~\ref{fig04}(b) 
shows that the average coupling follows an inverse power law
with respect to $\alpha$, as: 
$\sigma^{\rm eff}_{\rm asymptotic}(\alpha) =0.18\>  \alpha^{-1.073}$. 
This is consistent with the mean field approximation of Eq.~\eqref{eq01c},
where the fixed
point of the
Oja dynamics  indicates that $\left< \sigma_{jk}\right>_{t\to\infty}=1/\alpha$
and consequently $\sigma^{\rm eff}_{\rm asymptotic}=\sigma_c/\alpha = 0.2/\alpha$, see Eq.~\eqref{eq07}.

\par
\begin{figure}[H]
 \centering
\includegraphics[width=0.490\textwidth]{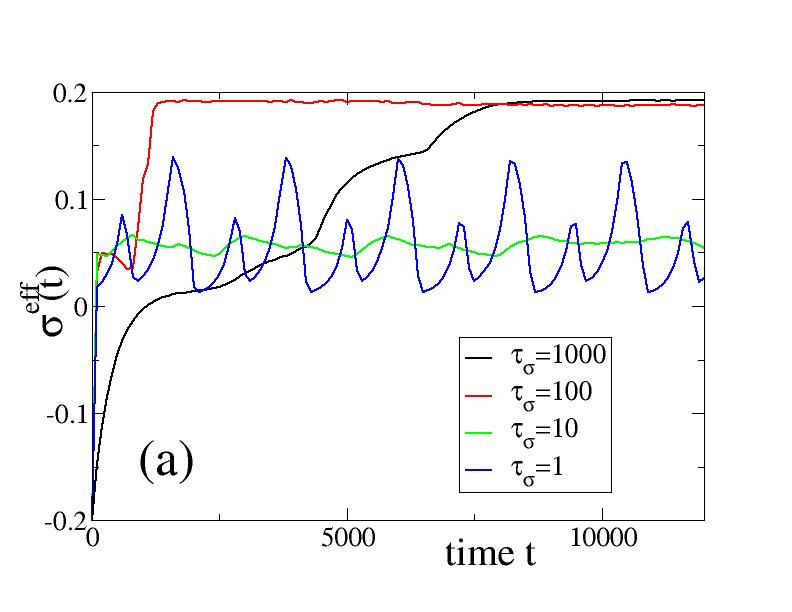}
\includegraphics[width=0.490\textwidth]{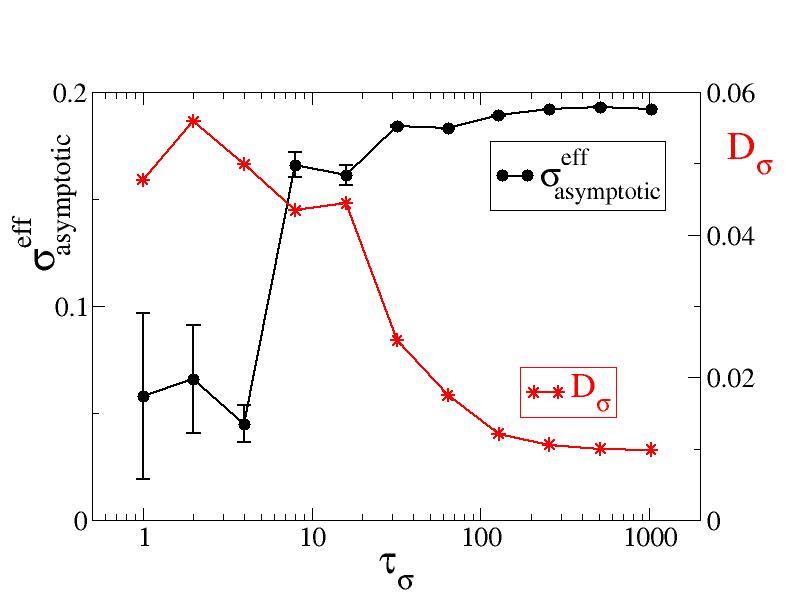}
 \caption{\label{fig05} FHN network with adaptive Hebb-Oja coupling and different time scales.
(a) The time evolution of the average coupling strength for  $\tau_{\sigma}=1$ (blue line),
10 (green line), 100 (red line) and 1000 (black line). (b) The asymptotic coupling 
$\sigma^{\rm eff}_{\rm asymptotic}$ (black line) and the spatial deviation $D_{\sigma}$ (red line) of the coupling
around the mean
as a function of the coupling time scale $\tau_{\sigma}$. The Oja parameter is set to 1 and all 
other parameter values are as in Fig.~\ref{fig01}.
}
\end{figure}

 Concerning the influence of the time scale $\tau_{\sigma}$ which governs the coupling 
evolution, in Fig.~\ref{fig05}(a) we present the system evolution at four different time scales:
$\tau_{\sigma}=1,\> 10,\> 100,\> 1000$. We note that for 
 the first two (low) time scales ($\tau_{\sigma}=1$ or 10) the system
reaches the same
asymptotics almost immediately, at about $t \sim 100-200$ TU. For the higher time scales, $(\tau_{\sigma}=100, \> 1000)$,
a different asymptotic state is reached at later times,
while the system goes through a number of structural transitions
as was discussed in the previous sections.
\par
\begin{figure}[H]
 \centering
\includegraphics[width=1.0\textwidth]{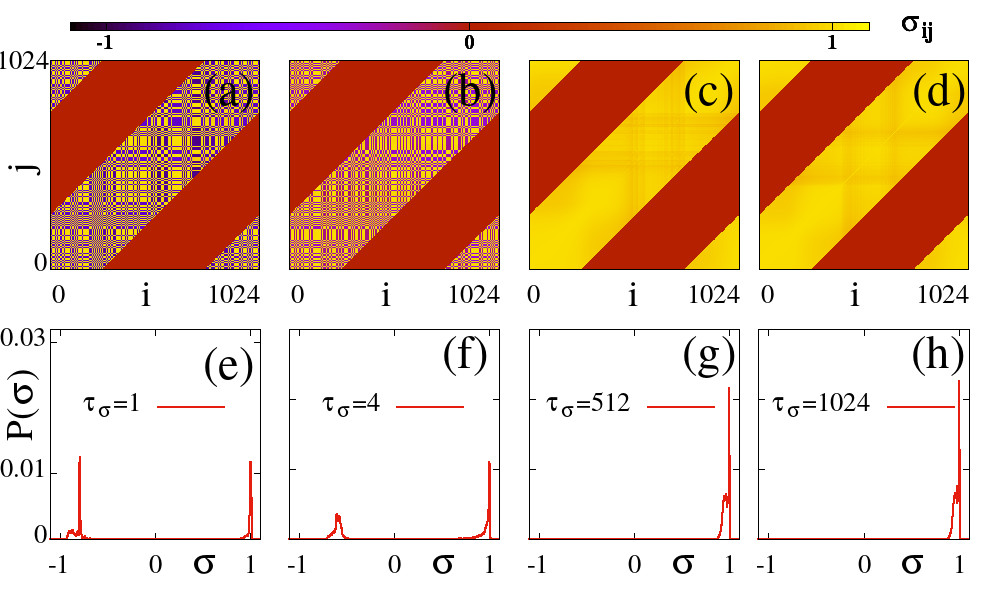}
\caption{\label{fig06} FHN network with adaptive Hebb-Oja coupling.
Top row: Spatial representations of the asymptotic couplings $\sigma_{ij}$ for:
(a) $\tau_{\sigma}=1$, (b) $\tau_{\sigma}=4$, (c) $\tau_{\sigma}=512$ and (d) $\tau_{\sigma}=1024$.
Bottom row: The corresponding histograms representing the distributions $P(\sigma)$ (probability
to find an effective coupling strength between $\sigma$ and $\sigma +\delta \sigma$).
Namely, the four histograms correspond to the four above cases as:
(e) $\tau_{\sigma}=1$, (f) $\tau_{\sigma}=4$, (g) $\tau_{\sigma}=512$ and (h) $\tau_{\sigma}=1024$.
We note that for low time scales $\tau_{\sigma}$ both negative and positive couplings survive
in the system, while for high $\tau_{\sigma}$ all coupling strengths accumulate in the 
positive axis.
 All parameter values are the same as in Fig.~\ref{fig01} and $\alpha =1$.
}
\end{figure}

  To determine the influence of the time scale parameter $\tau_{\sigma}$ on 
the asymptotics we plot in Fig.~\ref{fig05}(b) the asymptotic effective coupling strength, $\sigma^{\rm eff}_{\rm asymptotic}$, as a function of $\tau_{\sigma}$, see black circles. 
To cover short and long time scales we use a logarithmic $\tau_{\sigma}$ axis ($x$-axis).
Indeed, we note that for small time scales, e.g. $\tau_{\sigma} <  10$
for these parameter values, the average effective coupling strength
takes low values but with high temporal variations. The averages here are considered over time
and the variations are depicted by the errorbars. 
The temporal variations
in the low $\tau_{\sigma}$ scales are also clearly visible in Fig.~\ref{fig05}(a) for $\tau_{\sigma}=1$ and 10.
A transition takes place at $\tau_{\sigma}\approx 15$, and  $\sigma^{\rm eff}_{\rm asymptotic}$ shoots to high
values where a plateau is formed after about $\tau_{\sigma}\approx 250$. The temporal variations here are small,
and are not visible in the plot because the error bars
 are of the order of the symbol (circle) size. The decrease in the
size of the variations for large $\tau_{\sigma}$ scales is also visible from Fig.~\ref{fig05}(a), 
with $\tau_{\sigma}= 100$ and 1000. 
Apart from the temporal variations, in Fig.~\ref{fig05}(b) we also plot, in red color, the spatial
variations of the $\sigma_{ij}^{\rm eff}$ at the asymptotic limit. The calculations of $D_{\rm \sigma}$ follow
Eq.~\eqref{eq03}. As seen from the red curve in Fig.~\ref{fig05}(b), the spatial fluctuations are also large for
low $\tau_{\sigma}$ scales and after the abrupt transition they are reduced to minimal values.
A visual inspection of the
 spatial fluctuations in the coupling matrices $\sigma_{ij}$ is presented next, in  Fig.~\ref{fig06}.

\par In Fig.~\ref{fig06}, we have chosen to present the distribution of $\sigma_{ij}$
and not the distribution of $\sigma^{\rm eff}_{ij}$, in order to avoid complicated notation.
The reduction of $\sigma_{ij}$ to the effective values is trivial, since $\sigma^{\rm eff}_{ij}=\sigma_c\sigma_{ij}$
and $\sigma_c =$const(=0.2), as was discussed in Sec.~\ref{sec:network}.
At the top row of Fig.~\ref{fig06}, the asymptotic spatial structure of the coupling matrices  $\sigma_{ij}$
are presented for different values of $\tau_\sigma $, as indicated in the legend.
 In all four cases, the Oja parameter $\alpha =1$ and all other parameters are as in the working parameter set.
As initial conditions, the $\sigma_{ij}$ values were initiated randomly and uniformly 
around the value -1 in all cases.
\par
In particular, in Fig.~\ref{fig06}a we depict 
the asymptotic $\sigma_{ij}$-values for the case $\tau_\sigma =1$. 
The snapshot presented is at times $t=15000$ TU, where the system has reached its limiting state 
(see Fig.~\ref{fig05}a). 
The red-brown bands correspond to null coupling.
 From the image we can see that
 both positive (red-yellow)  and negative (black-blue) coupling values are present in the final coupling matrix. 
This can  be further verified from the corresponding histogram shown in Fig.~\ref{fig06}e, directly below Fig.~\ref{fig06}a, at the bottom row. 
When $\tau_{\sigma} $ is increased to 4, see Fig.~\ref{fig06}b and the corresponding histogram
Fig.~\ref{fig06}f, the 
asymptotic coupling matrix still contains positive (red-yellow)  and negative (black-blue) coupling values, but 
now the negative values are shifted to the right, toward 0. 

\par When  $\tau_{\sigma} $ is further increased to high scales, see Fig.~\ref{fig06}c for $\tau_{\sigma} =512$,
we note that the coupling strengths have moved to only positive values. This is more clearly visible from 
the corresponding histogram, Fig.~\ref{fig06}g. Slowing down the dynamics governing
the coupling evolution allows the coupling values to grow altogether
toward positive asymptotic values. Increasing further the time scale parameter, $\tau_\sigma$=1024, we
note that the coupling values still remain in the positive values and accumulate toward $\sigma_{ij}\to 1/\alpha=1$.

\par Overall, the time scale parameter $\tau_{\sigma}$ governing the dynamics of the coupling is highly influential
for the network evolution. Low $\tau_{\sigma}$ scales allowing for fast coupling evolution, cause 
partial confinement of the
coupling values in the negative domain resulting in heterogeneous coupling distributions. 
On the other hand, slow coupling evolution allows for smooth transitions of all links toward the asymptotic state
giving rise to a single, almost $\delta$-like, link-size distribution.

\subsection{Discussion on adaptation process and learning}
\label{sec:results-5}
\par 
In view of the above results,
we recall here that even in the non-adaptive FHN network there is a difference in the time scales
governing the dynamics of the membrane and the recovery potentials. Namely, the parameter $\epsilon$ 
in Eq.~\eqref{eq01a} shows that the membrane potentials $u_j$ evolve in time scales of the
order of $\epsilon^{-1}$. In the
present studies, mostly $\epsilon=0.01$, and this means that the membrane potentials evolve 100 times faster
than the recovery ones. In brain dynamics, it is known that neuron synapses undergo adaptive changes
in much larger times than the potential changes which are of the orders of milliseconds to seconds 
\cite{lazarevich:2017}.
 Synapses adaptation
drive processes such as learning (via gradual synapses growth or during
 structural plasticity in infancy and adolescence), 
or may
 decline due to aging or to the development of neurodegenerative disorders. These growth and decline adaptation processes
are known as ``long-term plasticity'', include processes such ``long-term potentiation'' and ``long-term depression''
\cite{morrison:2008,citri:2008} and take places in time scales much larger than milliseconds. 
The results in Secs.~\ref{sec:results-1} -~\ref{sec:results-4} indicate that  larger time scales $\tau_{\sigma}$ (here of the order of $10^3$) which lead to long term adaptive processes allow the network to explore many different
dynamical regimes before reaching its final steady state. 
These biologically-inspired long term adaptive
processes are behind the idea of
 using much larger time scales $\tau_{\sigma}$ during the integration
of Eq.~\eqref{eq01c}, with  $\tau_{\sigma}$ surpassing both the time scales of the membrane and the 
recovery potential evolutions. 
\par Along the same line and from an application-oriented perspective, these simulations provide useful 
insights for mechanisms related to neurodevelopmental processes, 
such as synapses' reinforcement (potentiation) and 
deterioration (pruning). In the  first case, as in Secs.~\ref{sec:results-1} and ~\ref{sec:results-2}, 
synapses increase their strength leading to learning occurring mainly
during early brain development. This can be assimilated by the present model, if at the initial stage
 $\sigma_{ij}$ values close to zero are considered, together with small (always positive) values of the Oja parameter.
For simplicity, the control parameter can be set to $\sigma_c=1$.
In this case, the $\sigma_{ij}$ values will gradually increase toward large values, asymptotically reaching $1/\alpha$,
imitating early-age learning processes (see Sec.~\ref{sec:quantitative}).  
In the second case, as in Sec.~\ref{sec:results-3}, the synapses' strengths decrease and
this can be considered as a result of aging or of neurodegenerative disorders. For the assimilation of
the second case, initially the $\sigma_{ij}$ values are set to large positive values while the Oja parameter
takes also large values. Starting with large $\sigma_{ij}$ at $t=0$, the coupling weights gradually deteriorate
reaching asymptotically values as small as $1/\alpha$. This process resembles the 
synapses' progressive deterioration leading to
the weakening of memory and other mental functions due to aging, Alzheimer, Parkinson 
or other neurodegenerative disorders.

%
%

\section{Conclusions}
\label{sec:conclusions}

In the present study, we investigate the influence of adaptive coupling in networks of 
interacting FitzHugh-Nagumo oscillators. We have chosen to explore the case of a ring network
with periodic boundary conditions. Such networks are known to produce chimera states in the
absence of adaptivity. In the present numerical investigations we use Hebbian network adaptivity
modified by the Oja forgetting term to prohibit the uncontrolled growth of the coupling weights.
The results indicate that if the time scales of the coupling are much slower than the 
time scales of the potential evolution, the network undergoes a number of transitions
through various synchronization states before reaching its limiting state. The intermediate
states may be synchronized states, or traveling waves or chimera states of various multiplicities.
The final state is largely determined by the values of the coupling control parameter, 
$\tau_{\sigma}$, and the Oja parameter, $\alpha$.
It is interesting to note that the transitions  between the different synchronization regimes,
as recorded by the average coupling strength mostly
and also by the Kuramoto order parameter, take place abruptly (non-smoothly).
We also show that the limiting average coupling strength decreases with the Oja parameter following
an inverse power law. With regards to the parameter controlling the adaptivity time scale, 
$\tau_{\sigma}$, we
report that the average coupling strength follows a lower plateau for small $\tau_{\sigma}$ and a
higher plateau for large $\tau_{\sigma}$ values. An abrupt transition takes place as the
parameter $\tau_{\sigma}$ crosses a turning point separating the two plateaus.
\par As earlier discussed, adaptive coupling is inspired by neural processes such as long- and short-term potentiation. The present results show that, in adaptive networks, coherence/incoherence transitions
are not achieved 
continuously but through a number of abrupt successive steps, in the case of large $\tau_{\sigma}$ scales.
Translated to the case of long-term potentiation, our results also indicate that learning  
does not proceed smoothly, but is achieved via a number of abrupt successive steps, 
at least in  systems/networks of finite size.

\par Apart from the present FHN model, adaptive transitions were earlier observed in the 
Leaky Integrate-and-Fire model \cite{provata:2025c} and could, therefore, be generic in neuron network systems. 
Regarding the Kuramoto model, preliminary studies
demonstrate that the gradual adjustment of the coupling variables via the Hebb-Oja rule 
do not lead to transitions between chimera
(or bump) states of different multiplicity because the coupling parameters (equivalent to $\sigma_{ij}$)
only mildly control the chimera multiplicity, which mostly depend on the coupling range $R$.\footnote{These
preliminary studies concern only the classical Kuramoto model, without delays
 or other modifications \cite{kuramoto:2002}.}
In the Kuramoto model,
 the different synchronization regimes may appear when the phase-lag parameter in the coupling term
(containing the $\sin$ expression) is varied \cite{oomelchenko:2018}. This parameter is not addressed
by the Hebb-Oja adaptation/learning rule.
\par Other open problems in this direction include the exploration of Hebb-Oja adaptivity in other networks
of nonlinear and neuronal oscillators, such as the
Hindmarsh-Rose, the van der Pol and the
Stuart-Landau oscillator. It would also be interesting to examine the persistence
of transitions at the limit of infinite network sizes and show if the transitions become
more abrupt as the networks grow to larger sizes.

\vskip 1cm
\noindent{\bf \large Acknowledgements} 

The authors would like to thank Drs. Yannis Almirantis and Wentian Li for helpful discussions.
This work was supported with computation time granted by the Greek Research \& Technology Network (GRNET)
in the National High Performance Computing HPC facility - ARIS - under Project ID PR016032.

\bibliographystyle{iopart-num}
\bibliography{./provata.bib}

\end{document}